\documentclass[12pt]{iopart}
\usepackage{iopams}
\usepackage{amssymb}
\usepackage{amstext}
\usepackage[english]{babel}
\usepackage[latin1]{inputenc}   % Enter your text in ISO-Latin 1
\usepackage[T1]{fontenc}        % To get scandinavian characters in the output
\usepackage{setspace}
\usepackage{float}
\usepackage{longtable}
\usepackage[dvips]{color}
\usepackage{graphicx}
\usepackage{multirow}
\usepackage{subfigure}
\usepackage{xspace}
\usepackage{rotating}
\usepackage{mathrsfs}
\usepackage{cleveref}
\usepackage{upgreek}
\usepackage{tcolorbox}
\usepackage[OT1]{fontenc}

\begin{document}

\title[]{Hydrogen redistribution in Zr-base  cladding  under
gradients in temperature and stress\footnote{\textbf{Presented in SMiRT-28, August 10-15, 2025, Toronto, Canada}.}}

\author{L. O.  Jernkvist \& A. R. Massih}

\address{Quantum Technologies AB, Uppsala Science Park, Uppsala SE-75183, Sweden}

\begin{abstract}
Computational models are  used here for simulating diffusion-controlled
redistribution of hydrogen that is picked up by zirconium-base nuclear
fuel cladding during light water reactor operation. Axial localization of
hydrogen, leading to localized precipitation of zirconium hydrides at
lower-temperature regions near interpellet gaps, is studied with a
bespoke model, while radial diffusion, leading to formation of a densely
hydrided rim subjacent to the waterside oxide layer, is studied with a
more general model. The calculated results are compared with
experimental observations and similar computational studies reported in
the literature. The results underline the importance of hydrogen
redistribution with regard to local embritt­lement of the cladding tubes.
\end{abstract}

\section{Introduction}
\label{sec:intro}

Zirconium base alloys utilized for fuel cladding in water-cooled nuclear
reactors pick up hydrogen as a consequence of oxidation and irradiation
during reactor service. When the hydrogen reaches its solid solubility
concentration in the metal ($\alpha$-phase), precipitates of zirconium
hydride ($\delta$-phase) nucleate and grow. The uptake of hydrogen and its
precipitation to hydride may potentially embrittle the cladding. The
embrittlement is generally localized to regions with low temperature
and/or high tensile stress, since hydrogen in solid solution
redistributes to these regions by migration along gradients in
temperature and stress. Light water reactor (LWR) fuel cladding
inevitably operates with radial, axial and circumferential temperature
gradients, which may be locally augmented by geometric irregularities,
e.g. interpellet gaps and spalled oxide layers.

To understand the localized embrittlement, modelling hydrogen redistribution in Zr-alloys
under temperature gradients has been an enduring effort over the years.
Early works include \cite{WAPD-TM-10},  \cite{sawatzky1961mathematics}, \cite{asher1970distribution} and \cite{marino1972numerical}. Many of these early models were not
suitable for treating conditions that prevail in LWRs, where the
cladding operate with temperature gradients while continuously picking
up hydrogen that precipitates into hydride phase. A model for hydrogen redistribution in the presence of radial and axial
temperature gradients in boiling water reactor (BWR) fuel cladding, suitable
for describing how hydrogen would localize in the interpellet region of
the cladding, was offered in \cite{Forsberg1990redistribution}. It used a single-phase description of hydride precipitation.

In the first part of our paper, we use an extended version of the \cite{Forsberg1990redistribution} method to compute
the time evolution of axial hydrogen redistribution both in $\alpha$-phase and
$\delta$-phase in a BWR Zircaloy-2 cladding close to an interpellet gap. In
the second part of the paper, we study the formation of a densely
hydrided rim, subjacent to the waterside oxide layer of severely
corroded pressurized water reactor (PWR) fuel cladding. Calculations are
done with a general model \cite{jernkvist2014multi}  that treats
temperature- and stress-induced hydrogen diffusion in both phases ($\alpha + \delta$),
in order to elucidate the effects of -hydride precipitates on hydrogen
transport.

\section{Single-phase thermal model}
\label{sec:single-phase}
The model used in the first part of our computations has been presented
previously  \cite{Forsberg1990redistribution} and in more detail in \cite{forsberg2025hydrogen} with some modifications. Here, we shall be brief.
\subsection{Governing equations}
\label{sec:goveq-1}
Hydrogen absorbed at the waterside surface of the zirconium alloy cladding tube diffuses into the material under temperature gradients with a flux $\mathbf{j}$. The basic equations describing the processes are
\begin{eqnarray}
\label{eqn:diffusion-eq}
\frac{\partial N}{\partial t} = -\nabla \cdot \mathbf{j}\\
\mathbf{j} = -D_\alpha\big(\nabla -Q^\ast_\alpha \nabla\beta\big)N_{s\alpha},
\label{eqn:Hflux-1}
\end{eqnarray}
where $\nabla$ is the gradient operator, $N$ is the total concentration of hydrogen, $N_{s\alpha}$ is the concentration of hydrogen in solid solution within the $\alpha$-phase metal, $D_\alpha$ and $Q^\ast_\alpha$ are the hydrogen diffusion coefficient and the heat of transport, respectively, in $\alpha$-phase, and $\beta=(RT)^{-1}$ is the inverse of temperature with $R$ being the molar gas constant. For $D_\alpha$ and $Q^\ast_\alpha$, we use the correlations presented in \cite{jernkvist2014multi}.
Equation \eref{eqn:Hflux-1} models the Soret effect \cite{deGroot2013non}, which causes migration of solute particles, here hydrogen atoms, from the hotter to colder region of the solid under a temperature gradient.

The above equations are solved with a quasi steady-state approximation: We know from the theory of diffusion that the mean distance travelled by the particle in a specimen during any time, $t_0$, is proportional to the square root of time, namely $\langle r^2 \rangle =4Dt_0$ in two dimensions, see e.g. \cite{Landau_Lifshitz_1959}. Using $D_\alpha$ for a typical cladding operating temperature of 583 K and a mean distance of 2 mm, $t_0 \approx 4$ hours. Since this time is short in comparison with the hydrogen uptake rate during reactor operation, the solution to eqs. \eref{eqn:diffusion-eq}-\eref{eqn:Hflux-1}  is approximately a sequence of equilibrium solutions. Hence, eq.  \eref{eqn:diffusion-eq}  in the model is replaced with a relation for the hydrogen uptake rate that defines the evolution of \emph{average} hydrogen concentration in the material, while a sequence of solutions to $\mathbf{j}=0$ provides the hydrogen \emph{distribution} at any time. As in  \cite{Forsberg1990redistribution}, in our calculations, we use a constant hydrogen uptake rate at the cladding outer surface, \emph{viz.}
\begin{equation}
\label{eqn:hprod}
\frac{dN_g}{dt} = c_1, \quad\quad  \Longrightarrow \quad\quad  N_g=c_0+c_1 t,
\end{equation}
where $c_0$ is the initial or residual hydrogen impurity concentration in the as-fabricated material, which is here set
to $c_0=0$, and $c_1$ is a constant that corresponds to a hydrogen uptake rate of 1062 $\mu$g(m$^2$day)$^{-1}$. This rate of hydrogen uptake may be considered as fairly high for Zr base cladding in LWR conditions.

As noted in the introduction, when hydrogen in zirconium alloys reaches a certain level of concentration at a given temperature, it precipitates into zirconium hydride (e.g. ZrH$_{1.6}$), referred to as the $\delta$-phase as distinguished from the $\alpha$-phase in which hydrogen resides in solid solution. The genre of this phase
transformation is nucleation with subsequent growth of nuclei to zirconium hydride platelets. The boundary of this phase transformation, i.e., temperature vs. concentration is expressed by
\begin{equation}
\label{eqn:tssp}
N_{p\alpha} = N_c \exp(-\beta H),
\end{equation}
where $N_{p\alpha}$ is referred to as the terminal solid solubility for hydride precipitation, or TSSP, $N_c$ is a constant and $H$ is the heat of mixing or the enthalpy of the $\delta$-phase formation. These quantities are determined by measurements which are affected by irradiation; see e.g. \cite{une2009terminal}. Numerical values for these
parameters, which we have used in our calculations, are listed in Table  \ref{tab:mod1-parameter}.
\begin{table}[htb]
  \caption{Constant model parameters used in calculations.}
   \vspace{-8mm}
  \label{tab:mod1-parameter}
  \begin{center}
      \begin{tabular}[h]{lll}
      \hline
    Symbol [unit]:      & Value     & Definition   \\
      \hline
     $N_c$ [wppm]& 32700 & Constant in eq. \eref{eqn:tssp}  for TSSP\\
     $\overline{H}$  [kJmol$^{-1}$]         &   25.04 & Mean value of $\delta$-phase formation enthalpy \\
      $\widetilde{\upsigma}_\mathrm{H}$  [kJmol$^{-1}$]        &  0.416 &  Standard deviation for  $\delta$-phase formation enthalpy  \\
      $N_\delta$ [wppm]       &  16500  & Hydrogen concentration in  $\delta$-phase at  $\delta/(\alpha+\delta)$ phase  boundary   \\
      \hline
      \end{tabular}
  \end{center}
\end{table}

\subsection{Methods of solution}
\label{sec:sol-1}
As discussed above, hydrogen uptake in Zr-base cladding under LWR conditions is much slower than hydrogen diffusion. This allows hydrogen to reach a state of equilibrium expediently with $\mathbf{j}=0$. So, in thermal equilibrium, eq. \eref{eqn:Hflux-1} can be integrated to give
\begin{equation}
\label{eqn:ss-sol}
N_{s\alpha} = C\exp(\beta Q^\ast_\alpha),
\end{equation}
where $C$ is an integration constant to be determined by the boundary condition. If the outer surface temperature of the cladding is considered to be constant, i.e. $T=T_o$ or $\beta = \beta_o$, corresponding to a concentration $N_u$ then
\begin{equation}
\label{eqn:ss-sol2}
N_{s\alpha} = N_u\exp[-(\beta_o-\beta) Q^\ast_\alpha].
\end{equation}
At the interphase boundary, $N_{s\alpha} = N_{p\alpha}$ and eqs. \eref{eqn:tssp} and \eref{eqn:ss-sol2} can be combined to give $H=\beta^{-1}\ln(N_c/N_{s\alpha})$. As in \cite{Forsberg1990redistribution}, we regard $H$ as a stochastic variable, subject to both thermal and material fluctuations. The former are due to equilibrium fluctuations that occur in any thermodynamic system as originally formulated by Einstein and well described in many classical texts, e.g. \cite{tolman1938principles}. The latter arise from inhomogeneities in the material, e.g., point defects, dislocations, grain boundaries, etc. and those caused by irradiation damage, which impact $H$ and thereby $N_{s\alpha}$.
What is more, when $\delta$-phase precipitates form and grow, local volume changes around the precipitates make fluctuations appreciable. Consequently, we take an approach inspired by the early work of Asher and Trowse \cite{asher1970distribution}, which suggested that the ``true'' hydrogen terminal solid solubility should be represented by a noisy line. As in Einstein's theory of fluctuation, we assume that the fluctuations in $H$ obey the Gaussian (normal) probability distribution. To that end, we define a dimensionless variable $x$ as
\begin{equation}
\label{eqn:enthalpy-dev-rat}
x = \frac{H - \mathbb{E}(H)}{\widetilde{\upsigma}_\mathrm{H}}\equiv \frac{\boldsymbol{\updelta}_H}{\widetilde{\upsigma}_\mathrm{H}},
\end{equation}
with the associated {probability density
\begin{equation}
\label{eqn:prob_density}
p(x) = (2\pi)^{-1/2} \exp(-x^2/2)\nonumber,
\end{equation}
where $\mathbb{E}(H)=\overline{H}$  is the \emph{expectation} (\emph{mean}) value of $H$ obtained from
measurement, and $\widetilde{\upsigma}_\mathrm{H}$ is the \emph{standard deviation} (square root of variance), which is a measure of fluctuations around the mean; see Table  \ref{tab:mod1-parameter}  for the values used. Then, at the $\alpha/(\alpha+\delta)$-phase boundary, eq. \eref{eqn:enthalpy-dev-rat} is expressed as
\begin{equation}
\label{eqn:enthalpy-dev-rat-boundary}
x = \frac{\beta^{-1}\ln(N_c/N_{s\alpha})- \mathbb{E}(H)}{\widetilde{\upsigma}_\mathrm{H}},
\end{equation}
As detailed in \cite{Forsberg1990redistribution}, upon computation of the probability distribution $ \mathcal{P}(x) $  the total hyd­rogen concentration in the $\alpha$- and $\delta$-phase at each $(r,z)$ position of the tube wall is calculated according to
\begin{eqnarray}
\label{eqn:tot-hyd}
 N(r,z) &=& \Big[1-\mathcal{P}(x)\Big] N_\alpha + \mathcal{P}(x) N_\delta,\\
 \mathcal{P}(x) &=& \int_{-\infty}^{x}p(t) dt.
\end{eqnarray}
Next, the cladding average hydrogen concentration is calculated by integrating eq. \eref{eqn:tot-hyd} radially and axially. The average value depends on the unknown parameter $N_u$ in eq. \eref{eqn:ss-sol2}, which can finally be determined from the conservation of hydrogen in the system: at any given time, the average hydrogen concentration should equate $N_g$, defined by the hydrogen uptake correlation in eq. \eref{eqn:hprod}. An iterative numerical solution for $N_u$ is used at each time step. Details of the method are presented in \cite{forsberg2025hydrogen} .

\subsection{Computations: Axial localization of hydrogen in BWR fuel
cladding}
\label{sec:compute-1}
In order to compute the hydrogen distribution locally in the cladding, we need as input to our model the temperature distribution in the cladding opposite the interpellet region, where there exists a discontinuity in pellet-cladding gap. The discontinuity studied here is caused by pellet chamfering, but in more severe cases, it may be aggravated by pellet stacking faults, missing pellet fragments, etc., which perturbs heat transfer across the gap. Local temperature calculations with chamfered pellets as shown in Fig. \ref{fig:temperature}a were done with a fuel rod modelling code. More precisely, a  $10\times10$ BWR UO\textsubscript{2} fuel bundle rod (Table  \ref{tab:roddat}), operating at a linear heat generation rate around 20 kWm$^{-1}$, was simulated. The temperature at the cladding outer surface was fixed at $T_o= 567$ K. The cladding wall temperatures were calculated at seven radial slices and fifteen axial nodes or segments, assuming symmetry with regard to the pellet midplane; see \cite{Forsberg1990redistribution}. We found it adequately accurate to use only the cladding inner surface temperature $T_i$ as an input and then compute the radial variation by a linear fit relation according to
\begin{equation}
\label{eqn:radtemp-prof}
T(r,z) = T_o +\Big[T_i(z)-T_o \Big]\Bigg(\frac{R_o-r}{R_o-R_i} \Bigg), \qquad  R_i\le r\le R_o,
\end{equation}
where $R_i$ and $R_o$ are the inner and outer radii of the cladding tube; see Table  \ref{tab:roddat}. Figure \ref{fig:temperature}a shows the input $T_i(z)$ and Fig. \ref{fig:H-vs-z}b shows the contour $T(r,z)$, calculated through eq. \eref{eqn:radtemp-prof} over 80 radial slices across the tube wall, as used in our numerical radial integrations.
\begin{table}[htb]
  \caption{Fuel rod (fuel pellet and cladding tube) design data used in calculations.}
    \vspace{-2mm}
  \label{tab:roddat}
  \begin{center}
      \begin{tabular}[h]{lll}
      \hline
     Design parameter  [unit]                    &  BWR $10 \times 10$     & PWR $17 \times 17$    \\
      \hline
     Fuel pellet material [--] & UO$_2$ & UO$_2$  \\
     Cladding material [--] & Zircaloy-4 & Std ZIRLO \\
      Pellet length    [mm]             & 10.0   & 9.83         \\
      Pellet diameter  [mm]             & 8.19 &  8.19     \\
      Pellet chamfer depth, $C_d$ [mm]        & 0.10 & 0.13      \\
      Pellet chamfer width, $C_w$ [mm]       & 0.25  & 0.51      \\
      Cladding outer radius, $R_0$ [mm]       & 4.810 & 4.750      \\
      Cladding inner radius, $R_i$   [mm]    &  4.180  &  4.178   \\
      \hline
      \end{tabular}
  \end{center}
\end{table}
\begin{figure}[htbp]
\begin{center}
\includegraphics[width=0.80\textwidth]{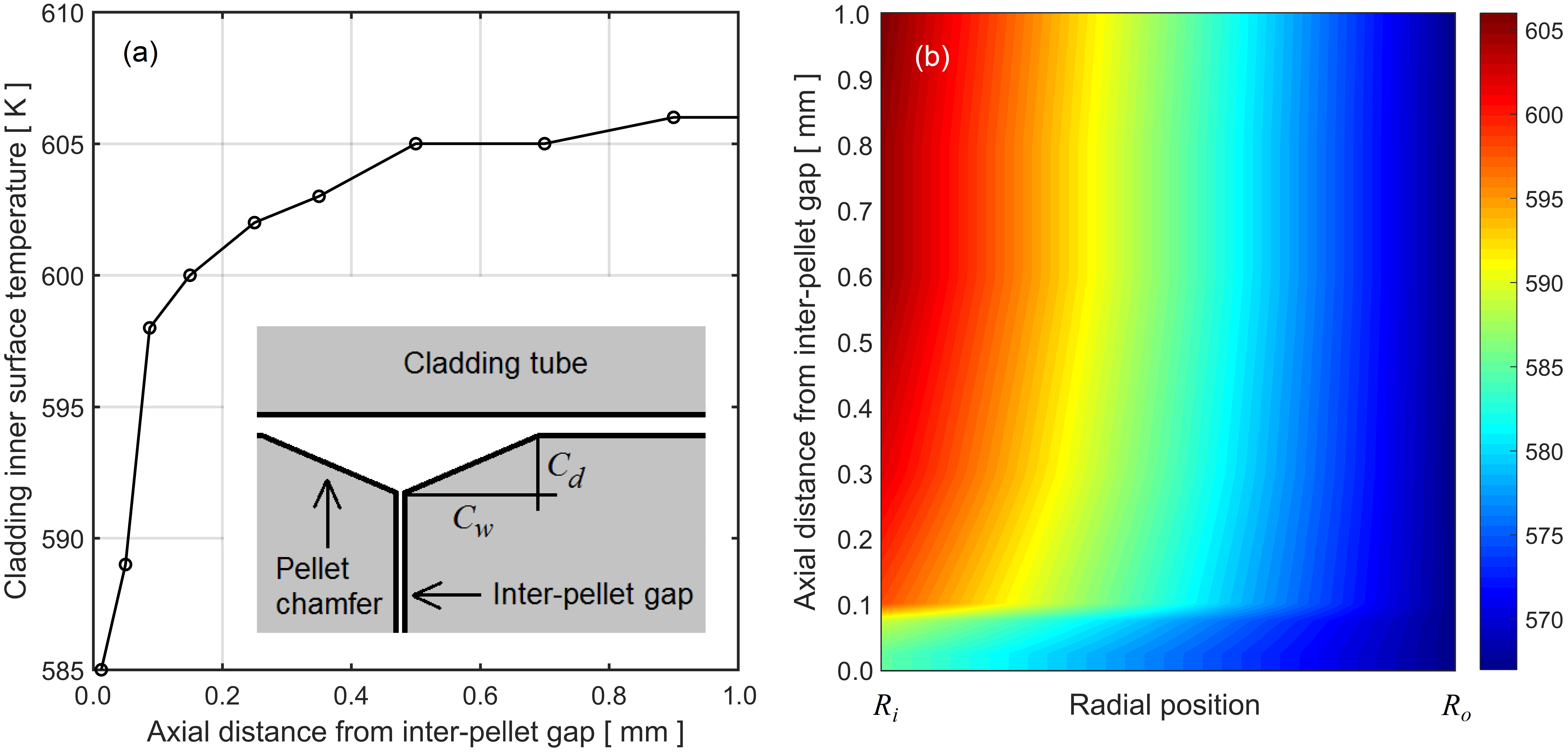}
\end{center}
\caption{(a) Cladding tube inner surface temperature
$T_i(z)$ close to the interpellet gap position at $z=0$. (b) Temperature distribution across the cladding, close to the interpellet gap.}
\label{fig:temperature}
\end{figure}
\begin{figure}[htbp]
\begin{center}
\includegraphics[width=0.85\textwidth]{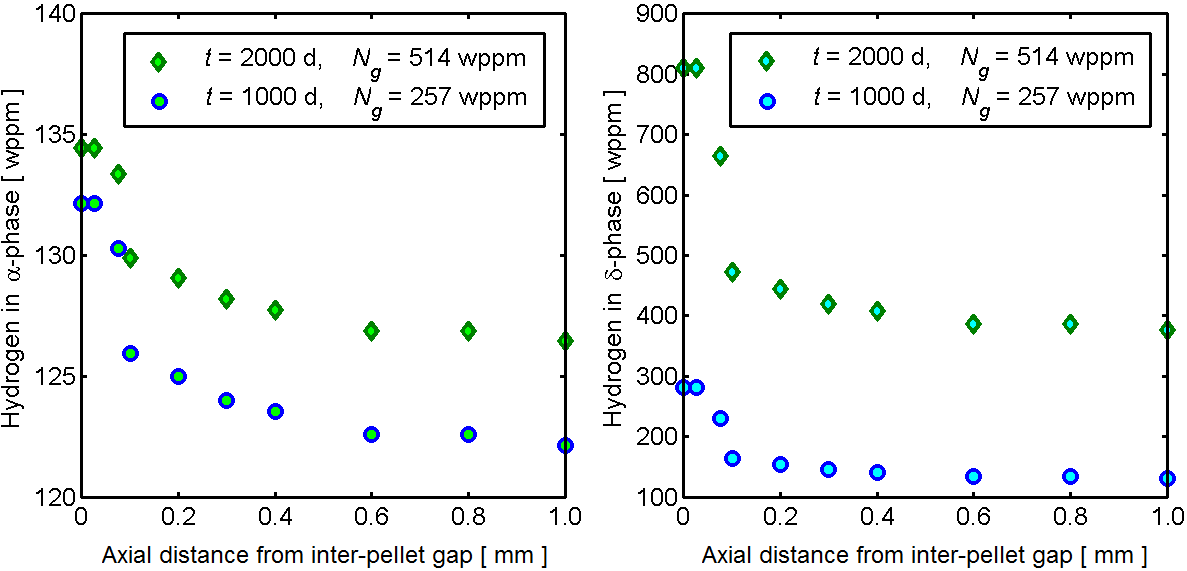}
\end{center}
\caption{Calculated wall-averaged cladding hydrogen concentrations in the vicinity of the interpellet gap at two stages of hydrogen uptake $N_g(t)$. Hydrogen in $\alpha$-phase (left) and $\delta$-phase (right).}
\label{fig:H-vs-z}
\end{figure}

Our model has been implemented in a dedicated computer program for calculation of hydrogen distribution in the cladding opposing the interpellet region. The program computes hydrogen located in solid solution ($\alpha$-phase) and in hydride precipitates ($\delta$-phase). The results of our computations for the $10\times10$ BWR fuel rod (Table   \ref{tab:roddat}) are depicted in Fig. \ref{fig:H-vs-z} at two occasions of hydrogen uptake, namely, after 1000 days and 2000 days, corresponding to the respective uptakes of 257 wppm and 514 wppm in the considered cladding segment.   Figure \ref{fig:H-vs-z}a shows hydrogen residing in the $\alpha$-phase, while  Fig. \ref{fig:H-vs-z}b shows the amount of hydrogen in the $\delta$-phase, averaged across the tube wall. A sharp increase in hydrogen concentration in $\delta$-phase after 2000 d is discernible opposite the interpellet gap. This behaviour is in qualitative agreement with few measured data reported in the open literature, see e.g. figure 10 in \cite{persson1991evaluation} and figure 1 in \cite{Forsberg1990redistribution}.

\section{Two-phase thermal-mechanical migration model}
\label{sec:two-phase}
The computational model presented above differs from conventional models for thermal diffusion of hydrogen in Zr alloys by considering fluctuations in the free energy of hydride formation, rather than using a deterministic terminal solid solubility for hydrogen. This evades a known problem with the conventional models, which unless rectified, they may yield singular solutions, characterized by excessive buildup of $\delta$-hydride in the coldest computational node(s). In particular, most models fail to reproduce the densely hydrided rim that forms beneath the oxide layer in severely corroded cladding. More precisely, they predict an infinitely narrow rim of practically pure hydride phase, corresponding to the outermost (coldest) compu­tational node. Various ad-hoc solutions have been proposed to rectify this shortcoming of conventional models.
For example,  Stafford \cite{stafford2015multidimensional} simply postulated an upper limit of 10 \% for the hydride volume fraction, Passelaigue et al. \cite{passelaigue2022predicting} empirically introduced hydrogen solubilities that depended on time and hydride volume fraction, McRae and Coleman \cite{mcrae2024heat} increased the heat of transport $Q^\ast_\alpha$ by a factor 4-5 as soon as hydrides formed in the metal, and Quecedo \cite{quecedo2024modelling} empirically reduced the hydride precipitation rate with increasing hydride volume fraction. Such ad-hoc measures are necessary to overcome inherent limitations of conventional single phase ($\alpha$-metal) hydrogen diffusion models. As shown below, no such measures are needed if diffusion of solute hydrogen is modelled in both the $\alpha$- and $\delta$-phase
material.

\subsection{Governing equations}
\label{sec:goveq-2}
The model applied below uses a continuum description of the two-phase ($\alpha$-metal + $\delta$-hydride) material. While eq. \eref{eqn:diffusion-eq} still holds, eq. \eref{eqn:Hflux-1} is replaced by
\begin{equation}
\label{eqn:conc-flux-gen}
\mathbf{j} =(1-\kappa)\mathbf{j}_\alpha + \kappa\mathbf{j}_\delta,
\end{equation}
where $\kappa$ is  the local volume fraction of $\delta$-hydride  and the hydrogen fluxes in the two phases ($i=\alpha,\delta$ ) are calculated separately
through an extended form of eq. \eref{eqn:Hflux-1} 
\begin{equation}
\label{eqn:conc-flux-alpha-delta}
\mathbf{j}_i = -D_i \Big(\nabla - Q_i^\ast \nabla \beta +\beta V_i^\ast  \nabla P_h\Big)N_{si}.
\end{equation}
Here, $V_i^\ast$ is the volume of transport, and $P_h$ is the hydrostatic pressure in the material. The latter is defined by $P_h=-\frac{1}{3}\mathrm{tr}(\boldsymbol{\sigma})$, where $\boldsymbol{\sigma}$ is the \emph{Cauchy stress tensor}. Hence, in contrast to the single-phase diffusion model described above, the two-phase model considers also
stress-induced hydrogen diffusion. Another difference is that the volume fraction of hydride, $\kappa$, is calculated through a deterministic point-kinetic phase transfor­mation model that considers hysteresis between hydride precipitation and dissolution.

In eq. \eref{eqn:conc-flux-alpha-delta}, $N_{si}$ are the concentrations of hydrogen in solid solution within the $\alpha$-metal and $\delta$-hydride phase, respectively.
$N_{s\delta}$ is introduced in order to account for hydrogen diffusion even in material consisting of pure hydride phase ($\kappa=1$). For reasons of local equilibrium at the metal-to-hydride interface, it is assumed that $N_{s\alpha}$  and $N_{s\delta}$ are comparable. More precisely, we postulate that $N_{s\delta}=cN_{s\alpha}$, where $c \in [0,1]$  is a dimensionless and constant model parameter that defines the jump of free hydrogen concentration at the metal-to-hydride interface. The model outlined above is fully described in \cite{jernkvist2014multi}. In particular, we show in Appendix D of \cite{jernkvist2014multi}, that when $c < 1$, there will always be a hydrogen flux downhill gradients in $\kappa$, i.e. away from densely hydrided regions. As will be shown below, this inherent feature of the two-phase model is important for reproducing hydride rim formation.

\subsection{Methods of solution}
\label{sec:sol-2}

In addition to the hydrogen transport equations described above, the applied model solves equations for time-dependent metal-hydride phase transformation, hydride orientation and mechanical equilibrium, including deformations and associated internal stresses caused by hydrogen and hydrides. The inter­connected equations are discretized in space by use of the finite element method, and implicit time stepping (\emph{backward Euler method}) is used for the discretization with respect to time. The discretized equations and their numerical solution are described in \cite{jernkvist2014multi}.

\subsection{Computations: Hydride rim formation in severely corroded
PWR fuel cladding}
\label{sec:compute-2}

To illustrate the model's capacity to reproduce the formation of a dense hydride rim, we will simulate the operating life of a PWR fuel rod of $17\times17$ design, irradiated to a rod average burnup of 68 MWd(kgU)$^{-1}$  in the Vandellos 2 reactor, Spain. The rod, which had standard ZIRLO material cladding, is thoroughly documented and analysed elsewhere, since a segment of it (known as CIP0-1) has been subjected to transient testing; see \cite{nea2022reactivity}, \cite{watanabe2005post} and \cite{georgenthum2017scanair}. This particular segment is studied here. Fundamental design parameters are given in Table  \ref{tab:roddat}.

Figure \ref{fig:lhgr-tclad} shows the rod average linear heat generation rate (ALHGR) versus time. Included in the figure are also local temperatures, calculated at
the cladding metal inner and outer surfaces, 2540 mm above the bottom of the fuel pellet column (BFC). The calculations were done with the FRAPCON-4.0P1 fuel rod performance program \cite{geelhood2015computer}. The calculated temperature histories in Fig. \ref{fig:lhgr-tclad}, together with FRAPCON's calculated histories for radial displacement of the cladding inner surface and corrosion-induced hydrogen flux across the outer surface, were used as thermal-mechanical boundary conditions for hydrogen transport calculations with the two-phase model. Hence, the operating life of the fuel rod was simulated with the aim to calculate the evolving radial distribution of hydrogen at the considered axial position. Axial symmetry was assumed, and a one-dimensional mesh with 50 equal-size quadratic elements were used. Only the cladding metal layer was modelled, meaning that the nodal radial spacing was 5 $\mu$m.
\begin{figure}[htbp]
\begin{center}
\includegraphics[width=0.8\textwidth]{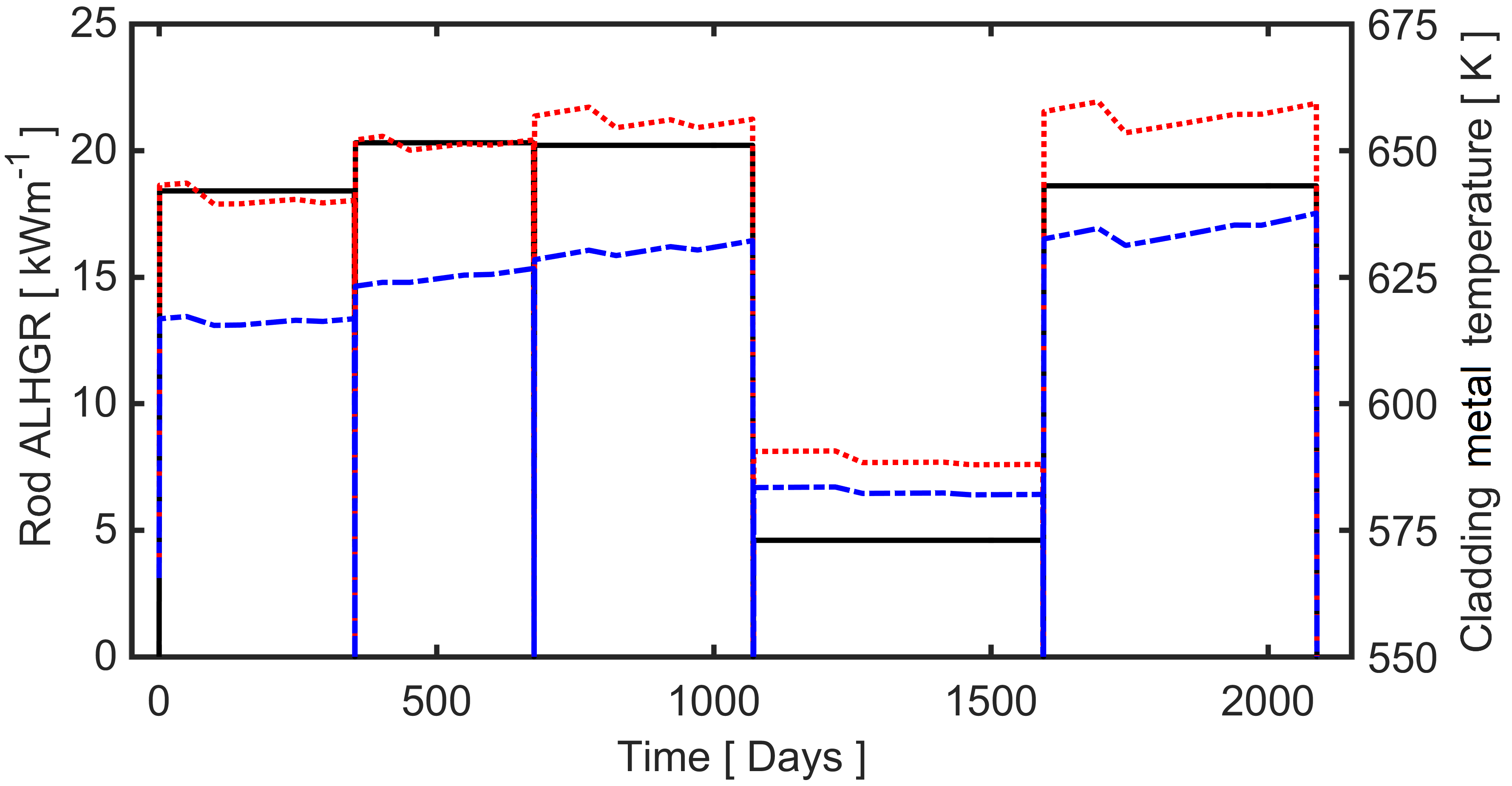}
\end{center}
 \vspace{-3mm}
\caption{ Rod average LHGR (solid black line) and calculated temperature histories for the cladding metal inner (red dotted line) and outer (blue dashed line) surface, 2540
mm above BFC. The cladding temperature drops to $<300$ K at zero power conditions.}
\label{fig:lhgr-tclad}
\end{figure}

Calculations were carried out with model parameters and material properties defined in \cite{jernkvist2014multi}, except for the parameter $c$ described above. More precisely, simulations were repeated with different values for $c$ to illustrate its importance for hydride rim formation. The results are shown in Fig. \ref{fig:combo-vs-r}a.
When the concentration of free hydrogen is postulated to be the same in $\alpha$- and $\delta$-phase ($c=1$), the model calculates an unrealistically high hydride concentration (14000 wppm) at the metal-oxide interface. By reducing the postulated ratio $c$, the peak value is reduced and the calculated hydride distribution becomes more uniform. The reader is referred to Appendix D in \cite{jernkvist2014multi} for a theoretical explanation to these results.
\begin{figure}[htbp]
\begin{center}
\includegraphics[width=0.85\textwidth]{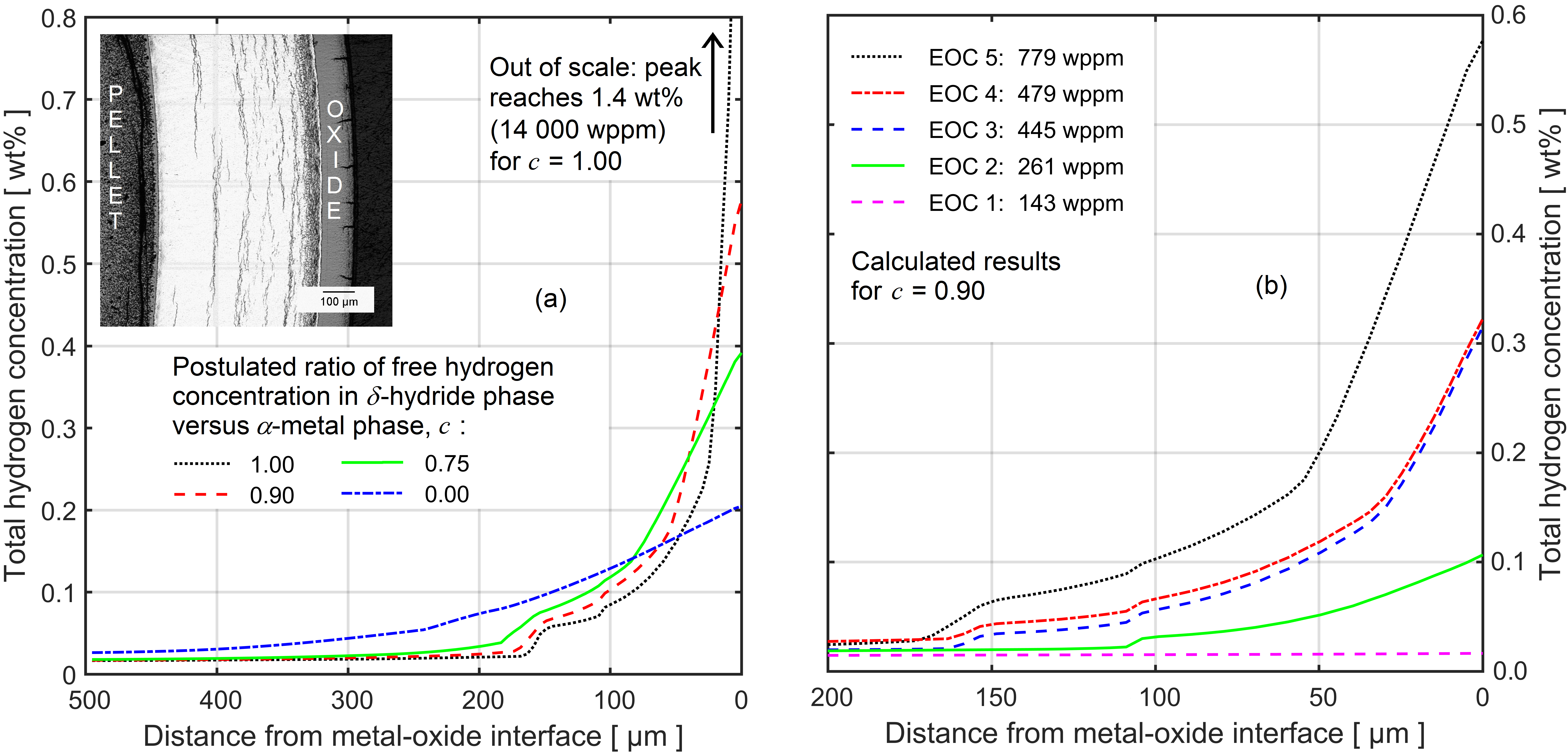}
\end{center}
\caption{ (a) Observed (insert) end-of-life hydride distribution, 2540 mm above BFC, in comparison with calculated results for different values of parameter $c$. (b) Calculated hydrogen distributions at end-of-cycle (EOC) 1 to 5. Calculated EOC wall-average hydrogen concentrations are indicated in the legend.}
\label{fig:combo-vs-r}
\end{figure}

The model in \cite{jernkvist2014multi} is one of few computational models that consider gradients in both stress and temperature as driving forces for hydrogen diffusion. Moreover, the model considers deformation of the material caused by free hydrogen (isotropic swelling) as well as hydride precipitation (anisotropic swelling). These deformations change the hydrostatic pressure within the material, and hence, they may affect the driving force for hydrogen diffusion; see eq. \eref{eqn:conc-flux-alpha-delta}. In particular, the hydrostatic pressure will increase in regions with hydride precipitation, thereby increasing the flux of free hydrogen away from these regions. This negative feedback effect suppresses local buildup of hydrides in the rim region, but the effect is weak. To illustrate this, we repeat the above calculations for $c=0.90$, but this time omitting any swelling induced by free hydrogen or hydride precipitates. In the nominal case, the swelling leads to a local volume increase at end of life of about 0.2 \% at the cladding inner surface, and to about 6 \% at the outer surface; see Fig. \ref{fig:hypothesis}a. This is a significant gradient that changes the stress distribution across the cladding wall: the red dotted line in Fig. \ref{fig:hypothesis}a shows the resulting change in local hydrostatic pressure ($\Delta P_h$), evaluated at end of life of the fuel rod. The radial gradient in hydrogen-induced swelling leads to a more tensile ($\Delta P_h<0$) stress state at the inner surface, while the stress state at the outer surface becomes more compressive ($\Delta P_h>0$). Since the hydride-induced part of the swelling is assumed to be anisotropic in our model, there is some stress relaxation by creep close to the metal-oxide interface: in our calculations, we assumed the oxide layer is cracked and imposes no mechanical restraint.

The calculated effect of the hydrogen-induced change in stress state on the end-of-life hydrogen distribution is shown in Fig. \ref{fig:hypothesis}b. Obviously, the change in stress distribution caused by non-uniform hydrogen- and hydride-induced swelling reduces the hydrogen concentration at the metal-oxide interface, while the hydride concentration in the inner part of the rim is increased. However, the calculated feedback effect from hydrogen-induced stresses is weak. This suggests that the solution to
eqs. \eref{eqn:ss-sol2}-\eref{eqn:tot-hyd} in the first part of our paper is adequate for capturing the steady-state hydrogen distribution.

\begin{figure}[htbp]
\begin{center}
\includegraphics[width=0.90\textwidth]{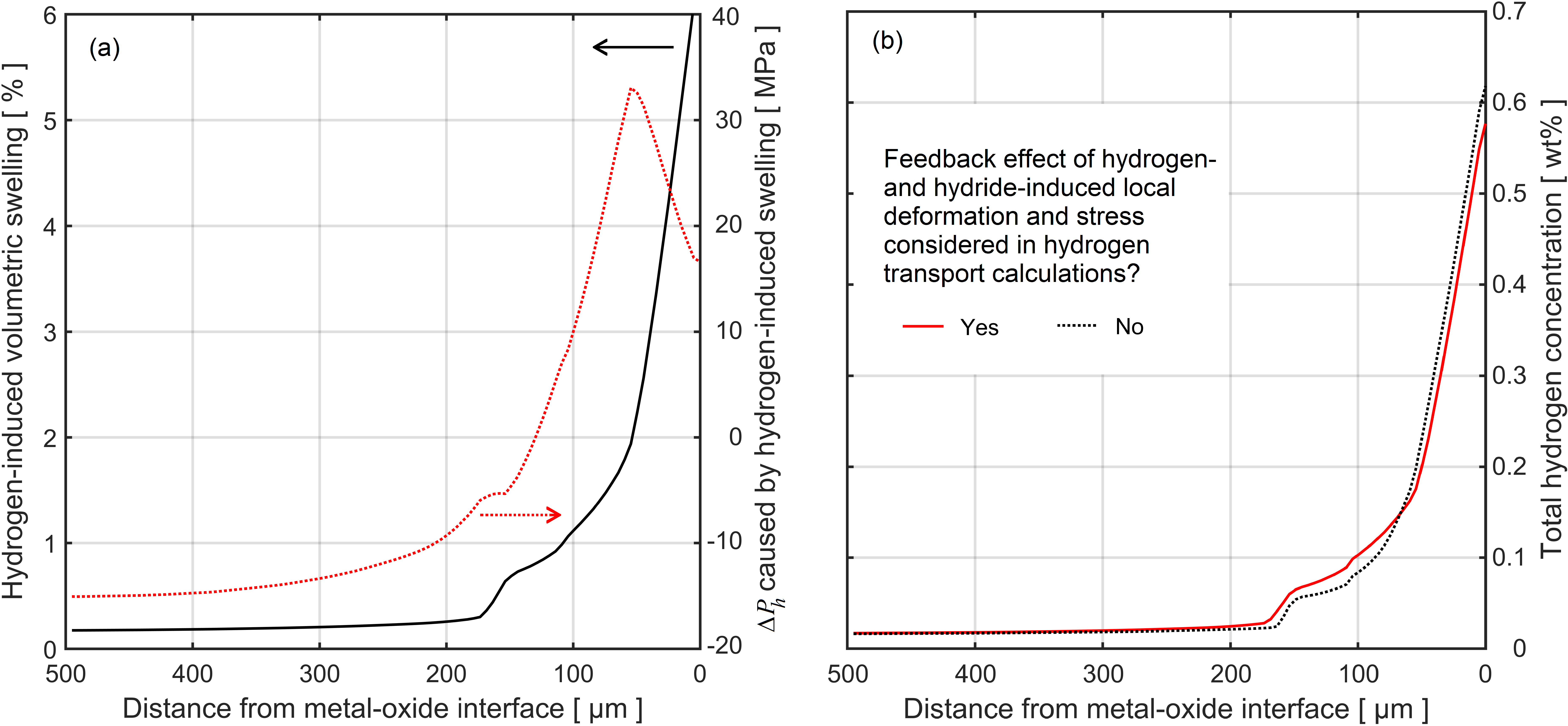}
\end{center}
 \vspace{-3mm}
\caption{(a): Calculated hydrogen-induced local swelling and resulting change in hydrostatic pressure. (b): Calculated feedback effect on end-of life hydrogen distribution
($c= 0.90$ in the calculations).}
\label{fig:hypothesis}
\end{figure}

\section{Conclusions}
\label{sec:conclude}

With a bespoke analytic method, we calculated hydrogen axial redistribution to lower-temperature regions, facing interpellet gaps, in a BWR Zircaloy-2 cladding tube. For an average hydrogen uptake of 514 wppm, we calculated an axial peak, radial average, total hydrogen concentration of 945 wppm. These results are in line with a similar computational study, performed on a PWR Zircaloy-4 rod by Nantes et al. \cite{nantes2024modeling}. Using a different model, but fairly similar thermal boundary conditions as in our study, they calculated an axial peak concentration of about 1150 wppm at an average hydrogen uptake of 474 wppm. These results underline the importance of hydrogen axial redistribution with regard to cladding local embrittlement.

We also simulated the formation of a densely hydrided rim, subjacent to the waterside oxide layer of severely corroded PWR fuel ZIRLO cladding. For these analyses, we used a more general model, which considered temperature- and stress-induced diffusion of solute hydrogen in both $\alpha$-phase metal and $\delta$-phase hydride. The model gave reasonable results with regard to the radial width and hydride density of the rim, if the solute hydrogen concentration in the $\delta$-phase was assumed to be 50-90\% of that in the $\alpha$-metal. Moreover, the calculated hydrogen diffusion was only weakly affected by feedback effects from changes in the stress field, caused by local hydrogen- and hydride-induced swelling.

Finally, the two models presented here use different approaches for avoiding singular and unphysical results with excessive buildup of hydride in the coldest computational node. Both approaches are physically based, in contrast to many ad-hoc solutions proposed in the literature.

\section*{Acknowledgments}
This work was supported by the Swedish Radiation Safety Authority
through grant SSM 2023-4412.

\section*{References}
\bibliographystyle{alpha}
\bibliography{microRev2}

\end{document}